\documentclass[english,pra,aps,twocolumn,showpacs]{revtex4}
\usepackage[T1]{fontenc}
\usepackage[latin1]{inputenc}
\usepackage{babel,amsmath,graphicx}

\newcommand \beq{\begin{equation}}
\newcommand \beqa{\begin{eqnarray}}
\newcommand \beqann{\begin{eqnarray*}}
\newcommand \eeq{\end{equation}}
\newcommand \eeqa{\end{eqnarray}}
\newcommand \eeqann{\end{eqnarray*}}


\begin{document}

\title{Energy-dependent effective interactions for dilute
many-body systems}
\author{A. Collin}
\author{P. Massignan}
\altaffiliation[Present address: ]{Institute for Theoretical Physics, University of Utrecht, Leuvenlaan 4, 3584 CE Utrecht, The Netherlands.}
\author{C. J. Pethick}
\affiliation{NORDITA, Blegdamsvej 17, DK-2100 Copenhagen~\O, Denmark}
\begin{abstract}

We address the issue of determining an effective two-body interaction
for mean-field calculations of energies of many-body
systems.  We show that the effective interaction is proportional to the
phase shift, and demonstrate this result in the quasiclassical
approximation when there is a trapping potential in addition to the
short-range interaction between a pair of particles.  We calculate
numerically energy levels for the case of an interaction with a
short-range
square-well and a harmonic trapping potential and show that the
numerical results agree well with the analytical expression.  We derive
a generalized Gross--Pitaevskii equation which includes effective range
corrections and discuss the form of the electron--atom effective
interaction to be used in calculations of Rydberg atoms and molecules.

\end{abstract}
\date{\today}

\maketitle

\section{Introduction}

Ultracold gases represent an ideal environment to study fundamental
processes such as interparticle collisions and molecule formation.
These systems are generally dilute, in the sense that the mean distance
between particles $\sim n^{-1/3}$ is typically larger than the range $R$
of the interactions:  under these conditions many-body encounters are
rare, and interactions can be satisfactorily modeled by two-body
collisions.

Atomic interaction potentials have a complicated structure and are
generally not known exactly but, as shown by Fermi \cite{Fermi36}, the
long-wavelength, low-energy properties of the two-body system can be
reproduced exactly if one replaces the true potential by a suitable
boundary condition for the relative wave function at the origin,
\begin{equation}
\psi(r)\propto 1-\frac{a}{r},
\end{equation}
where $\bf r$ is the relative coordinate vector.  This boundary
condition depends on a single parameter, the scattering length
$a=-\lim_{k\rightarrow 0}\delta/k$, where $\delta$ is the s-wave
scattering phase shift. When employed in the two-body problem, this
leads to an energy shift given by
\begin{equation}
\label{ScLengthDeltaE}
\Delta E=\frac{2 \pi \hbar^2 a}{\mu}|\psi(0)|^2,
\end{equation}
where $\psi({\bf r})$ is the relative wave function in the absence of
two-body interactions and $\mu$ is the reduced mass of the two
particles. This result leads to the interaction term in the
Gross--Pitaevskii
(GP) equation for the mean-field wave function of a Bose--Einstein
condensed gas.  The standard Fermi treatment is justified as long as the relative momentum
$k$ is so low that $k|a|\ll 1$, but to deal with phenomena at higher
energies the theory must be improved.

An extension of the Fermi pseudopotential was proposed in Refs.\
\cite{Huang,HuangCorr}, where the authors introduced a more general
contact pseudopotential which depends on energy and which includes all
partial waves $l$.  Its s-wave component is given by
\beq
  V_{\rm ps} \psi =  - \frac{2\pi\hbar^2}{\mu}
\frac{\tan\delta}{k}\delta({\bf r})\frac{\partial r\psi}{\partial
r} . \eeq The solution of the Schr\"odinger equation for the
pseudopotential agrees with that for the actual potential in the
region where the actual potential vanishes.  The magnitude of the
pseudopotential is specified in terms of the tangent of the
phase shift, $\tan \delta(k)$, where the wave number $k$ must
be taken to be that in the absence of the potential.
As shown in
\cite{Blume02,Bolda02,Idziaszek06}, when the Schr\"odinger equation is
solved for this pseudopotential
one obtains eigenenergies and wave functions that reproduce very
accurately the results obtained from integrating the Schr\"odinger
equation directly for the actual microscopic potential. Following
the suggestion of Ref.\ \cite{Omont77}, such a pseudopotential has
been employed in studies of \mbox{Rydberg} atoms and molecules
\cite{Greene00, Hamilton02}.  However, when used in mean-field
calculations of the energy, it yields divergent energy shifts when
a phase shift becomes close to $\pi/2$ (modulo $\pi$). A further
proposal for the relationship between an effective interaction and the
phase shift has been made in the
context of deriving a generalization of the GP equation, where it has
been argued that the energy shift should be  proportional to the
real part of the forward scattering amplitude, i.e.\ $\Delta E\propto \sin \delta
\cos \delta$ \cite{Gao03}.

In this paper we explore the relationship between energy shifts and
phase shifts.  We shall argue that the generalization of Eq.\ (\ref{ScLengthDeltaE})
 to nonzero $k$ is to replace the scattering length by
$-\delta/k$, a result demonstrated long ago for particles in the absence
of a trapping potential \cite{Fumi55}.  In Sec.\ II we first demonstrate
this result for two particles whose relative motion is confined to lie
within a sphere, and then show that in the quasiclassical approximation
it also holds if there is an additional trapping potential that depends
on the relative coordinate.  Following that we calculate numerically the
energy shift for an interaction consisting of a short-range square well
plus a harmonic trapping potential and show that the analytical formula
fits the data well.  In Sec.\ \ref{subsec:EffRange}, we derive a
generalization of the GP equation that includes the effective range of
the interaction through a derivative term and compare results with
predictions based on the prescription of Ref.\ \cite{Gao03}.  In
Sec.~\ref{sec:Rydb} we analyze effective interactions for ultracold
\mbox{Rydberg} atoms and molecules.  We summarize our results in
Sec.~\ref{sec:conc}.

\section{Energy shift}
\label{sec:EnergyShift}

The relationship between energy shifts and phase shifts for a
particle interacting with a static impurity was analyzed long ago
(see e.g.,~\cite{Fumi55}), and we briefly review the argument in the
context of the two-particle problem.
Consider
two particles, with reduced mass $\mu$, interacting via a
spherically-symmetric potential.
In the absence of interaction, the relative wave function
for an s-state is of the form
\begin{equation}
\psi(r)=A\frac{\sin(k_{0}r)}{r}.
\label{psifree}
\end{equation}
For definiteness, we imagine the relative motion to be confined by a
sphere of radius $R$, and we impose the boundary condition that the wave
function must vanish
at $r=R$.  This implies that
the wave number in the absence of interaction is $k_{0}R=n\pi$, and
normalization of the wave function gives
$A=(2\pi R)^{-1/2}$.
In the presence of a short-ranged interaction $V_\mathrm{sr}(r)$, which we shall assume vanishes
more rapidly than $r^{-1}$ for large $r$,
the asymptotic wave function
will have the same form with a phase shift:
\begin{equation}
\psi(r)=A\frac{\sin(kr+\delta)}{r}.
\end{equation}
To satisfy the boundary condition at $r=R$, the wave number must now
obey the equation
$kR+\delta=n\pi$,  which implies that the
wave vector is shifted by an amount $\Delta k=k-k_{0}=-\delta/R$. The
energy shift is then given by
\begin{equation}
\label{BOXDeltaE}
\Delta E\simeq\frac{\hbar^{2}}{\mu}k_0\Delta k=
\frac{2\pi\hbar^{2}}{\mu}\left(-\frac{\delta}{k_0}\right)\left|\psi(0)\right|^{2}.
\end{equation}
Thus the energy shift due to interparticle interaction is proportional
to the phase shift $\delta$.  In the limit of zero energy,
scattering theory (see e.g.,~\cite{LandauQMechBook}) shows that the
s-wave phase shift is proportional to the wave vector $\delta \simeq
-ka$ and one recovers the well-known result that the effective
interaction has a contact form, with strength
$U_0=2\pi\hbar^{2}a/\mu$.

\subsection{Presence of an external confining potential}

One may ask whether the result (\ref{BOXDeltaE}) applies in the
presence of a trapping potential.  We therefore consider the same
problem as above, but with an additional external potential
$V_{\rm ex}(r)$ for the relative motion \cite{harmonic}.  We shall
assume that $V_{\rm ex}(r)$ increases with increasing $r$,
and we shall impose the boundary condition that the wave function
tends to zero for large $r$.

The basic assumption we shall make is that the trapping potential $V_{\rm ex}(r)$ varies negligibly over both the range $L$ of the two-body potential $V_{\rm sr}(r)$ and over
length scales $\sim |\delta/k|$, which we shall show will play the role
of
an energy-dependent scattering length. In addition, we assume that $V_{\rm ex}(r)$ varies sufficiently slowly in
space and that the energy of the state is sufficiently high that
we may employ the quasiclassical approximation. As usual, it is
convenient to work
with the quantity $\chi=r\psi$, and in terms of it, the  Schr\"odinger
equation becomes \beq
-\frac{\hbar^2}{2\mu}\frac{d^2\chi}{dr^2}+V\chi=E\chi, \eeq and we
may take $\chi$ to be real.
The quasiclassical approximation
for the potential  $V_{\rm ex}(r)$ is
\begin{equation}
\label{WKBwavefunction}
\psi(r)=\frac {A} {r \sqrt{p(r)}} \sin\left[\int_0^r  dr'\sqrt{\frac {2\mu[E-V_{\rm ex}(r')]}{\hbar^2}}\right],
\end{equation}
where $p(r)=\sqrt{2\mu[E-V_{\rm ex}(r)]}$ is the classical relative momentum of
the two particles.  The normalization constant $A$ is fixed by requiring
that the norm of $\psi$, i.e.\ the volume integral of $|\psi|^2$
out to the classical turning
point
$r=r_c$ where
$V_{\rm ex}(r_{\rm c})=E$, be unity.  Since in Eq.\
(\ref{WKBwavefunction})
$p^{-1/2}$ changes
slowly over a period of oscillation of the sine function, we obtain
\begin{equation}
\label{WKBnormalization}
\int_0^{r_{\rm c}}d\mathbf{r}\ \psi^2(r)=2\pi A^2 \int_0^{r_{\rm c}}
\frac{dr}{p(r)}=1.
\end{equation}
By choosing the zero of energy such that
$V_{\rm ex}(0)=0$, the amplitude
of the wave function when the positions of the two particles coincide is
found to be
\begin{equation}
\label{WKBorigin}
\psi(0)=\frac{\sqrt[4]{2\mu E/\hbar^4}}{\sqrt{2\pi\int_0^{r_{\rm c}}dr\
p^{-1}(r)}}.
\end{equation}

The phase of the semi-classical wave function obeys Bohr's quantization rule \cite{LandauQMechBook}
\begin{equation}
\label{BohrQuantization0}
\int_0^{r_{\rm c}} dr\ \sqrt{\frac
{2\mu[E-V_{\rm ex}(r)]}{\hbar^2}}=(n+\alpha)\pi,
\end{equation}
where $n$ is a positive integer and $\alpha$ is a constant that depends on the nature of the potential
in the vicinity of the classical turning point. The presence of
the potential $V_\mathrm{sr}(r)$ induces an
energy shift $\Delta E$, and asymptotically the wave function acquires a phase shift $\delta$ that
satisfies
\begin{equation}
\label{BohrQuantization}
\int_0^{r_{\rm c}'} dr\ \sqrt{\frac {2\mu[E+\Delta
E-V_{\rm ex}(r)]}{\hbar^2}}+\delta(E+\Delta E)=(n+\alpha)\pi,
\end{equation}
where $r_{\rm c}'$ is the appropriate classical turning point in the
presence of the short-range interaction.
By taking the difference between Eqs.\ (\ref{BohrQuantization}) and (\ref{BohrQuantization0}),
expanding the integral to first order in $\Delta E$, and
making use of Eq.~(\ref{WKBorigin}), one obtains the result
\begin{equation}
\label{WKBDeltaE}
\Delta E=-\frac{2 \pi \hbar^2}{\mu}\frac{\delta(E+\Delta E)}{\sqrt{2 \mu E/\hbar^2}}|\psi(0)|^2,
\end{equation}
in agreement with Eq.~(\ref{BOXDeltaE}).  The difference between $r_c$
and $r_{\rm c}'$ plays no role since the integrand vanishes at the
turning point.

\subsection{A numerical example}
We now perform numerical calculations of the energy shift for a
model
potential. For the trapping potential, we choose a harmonic potential of frequency $\omega$,
$V_{\rm ex}(r)=\mu\omega^2r^2/2$, while for the particle-particle  interaction we
take an attractive spherically-symmetric potential which is equal to a constant,
$V_0<0$, for $r<L$ and zero otherwise. The problem is similar to that
considered in Refs.\ \cite{Bolda02, Blume02,Idziaszek06}, but with a
simplified short-range interaction.
For $r<L$, the solution with energy $E$ that is regular at the origin is
given by Eq.\ (\ref{psifree}) with
wave vector $k=\sqrt{2\mu(E-V_0)/\hbar^2}$,
\begin{equation}
\label{psiIn}
\psi_{\mathrm{in}}(r)\propto \frac{\sin(kr)}{r},
\end{equation}
For $r>L$, the wave function is the general solution of the
Schr\"odinger equation for the harmonic potential,
\begin{multline}
\label{psiOut}
\psi_{\mathrm{out}}(r)\propto e^{-r^2/2 a_{\mathrm{ho}}^2} \frac{\sqrt{\eta }}{a_{\mathrm{ho}}} \left[\, _1F_1\left(\frac{3-\eta }{4},\frac{3}{2},\frac{r^2}{a_{\mathrm{ho}}^2}\right)\right.\\
\left.-C\frac{a_{\mathrm{ho}}}{r}\, _1F_1\left(\frac{1-\eta
}{4},\frac{1}{2},\frac{r^2}{a_{\mathrm{ho}}^2}\right)\right],
\end{multline}
where $\eta=2E/\hbar\omega$, $\, _1F_1(\alpha,\gamma,z)$ is the confluent hypergeometric
function of the first kind \cite{Abramowitz} and the constant
$C=\Gamma \left(\frac{1-\eta}{4}\right)/2 \Gamma \left(\frac{3-\eta }{4}\right)$
is chosen to ensure that $\psi_{\mathrm{out}}(r)$ vanishes at infinity \cite{Busch98,Jonsell02}.
The allowed energies are found by matching at the boundary $r=L$ the
logarithmic derivatives of the solutions inside and outside the well.

The phase shift $\delta$ for a given energy $E$ may be found by equating
at
$r=L$ the logarithmic derivatives of $\psi_{\mathrm{in}}$ and of the
solution outside the core in the absence of the smooth potential, i.e.\ a
spherical wave with wave vector $k_0=\sqrt{2\mu E/\hbar^2}$.  This
yields the relation

\begin{equation}
\label{deltaE}
\delta(E)=\arctan\left[\frac{k_0}{k}\tan(k L)\right]-k_0 L.
\end{equation}

\begin{figure}

\centerline{\includegraphics[width=8.6cm,clip=]{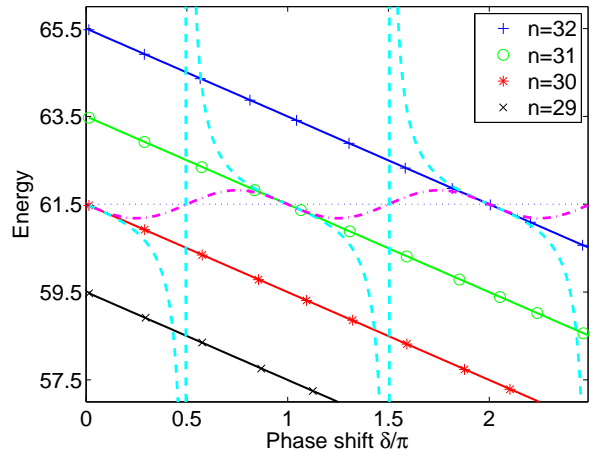}}

\caption{Energy levels for the model potential described in the text as
a function of phase shift, which is given by Eq.\
(\ref{deltaE}). Levels are indexed by the principal
quantum number $n$.  The symbols indicate the results of
numerically solving
the matching condition for the Schr\"odinger equation at $r=L$.
The values of the energies
$E(V_\mathrm{sr})=E(0)+\Delta E(V_\mathrm{sr})$ calculated using Eq.\
(\ref{WKBDeltaE}) (continuous lines) are indistinguishable from the
 values (symbols).  As a comparison, we also show the energy shifts
given by Eq.\ (\ref{GreenesDeltaE}) (dashed line) and Eq.\
(\ref{GaosDeltaE}) (dashed-dotted line).  The energy is measured in units of
$\hbar \omega$ and the core width is $L=2a_{\mathrm{ho}}$.}

\label{fig:DeltaEvsDelta}
\end{figure}

In Fig.\ \ref{fig:DeltaEvsDelta} we show results for the energy of a few
states with many nodes, obtained by direct calculation. The energy levels of
our model are closely reproduced by Eq.\ (\ref{BOXDeltaE}), while the predictions given in Ref.\ \cite{Greene00},
\begin{equation}
\label{GreenesDeltaE}
\Delta E=-\frac{2 \pi
\hbar^2}{\mu}\left[\frac{\tan\delta}{k}\right]|\psi(0)|^2,
\end{equation}
and in Ref.\ \cite{Gao03},
\begin{equation}
\label{GaosDeltaE}
\Delta E=-\frac{2 \pi
\hbar^2}{\mu}\left[\frac{\sin\delta\cos\delta}{k}\right]|\psi(0)|^2,
\end{equation}
are  correct only for small values of $|\delta|$ (modulo $\pi$).
Note that if $|\delta|>\pi/2$ and $|\delta/\pi-n|\ll 1$ (where $n$ is a suitable integer),
the theoretical values of the energy given by expressions (\ref{GreenesDeltaE})  and  (\ref{GaosDeltaE}) agree with an energy eigenvalue of
the system, but one with a different number of nodes inside the
short-range potential.

\section{A generalized Gross--Pitaevskii equation}
\label{subsec:EffRange}

In the numerical example above, we considered a state with many nodes.
For applications to Bose--Einstein condensates, the relevant wave
numbers
are usually small, and therefore it is interesting to look at the opposite case of small but nonzero
wave numbers. When the Wigner threshold condition
$k|a|\ll
1$ is violated, the s-wave phase shift is no longer
linear in the wave vector, and its energy dependence can be written for
potentials that fall off more rapidly than $r^{-5}$ for large $r$ as
\begin{equation}
\label{effectiverange}
k\cot\delta=-\frac{1}{a}+\frac{1}{2}r_{\rm e}k^{2}+o\left(k^{2}\right),
\end{equation}
where $r_{\rm e}$ is the effective range of the interaction \cite{breakdown}. For small phase
shifts $\delta \simeq \tan\delta-\tan^{3}\delta/3$ and
we find:\begin{equation}
-\frac{\delta}{k}=a\left(1-g_{2}k^{2}\right)\label{eq:MinusDeltaOverKWithG2}\end{equation}
where we have introduced
\begin{equation}
g_{2}=\frac{a^{2}}{3}-\frac{ar_{\rm e}}{2}.
\label{g2}
\end{equation}
The energy shift is then given by
\begin{equation}
\label{g2DeltaE}
\Delta E=\frac{2\pi\hbar^{2}a}{\mu}\left[1-g_{2}k^{2}\right]\left|\psi(0)\right|^{2}.
\end{equation}

It is interesting to note that in the case of hard spheres of diameter
$a$ the boundary condition on the surface of the sphere implies that
$\delta=-ka$ for all $k$ and therefore $g_2$ (and all higher terms in an
expansion of $\delta$ in powers of $k$) should vanish. Since from Eq.\ (\ref{effectiverange}) $r_{\rm e}=2a/3$ for the hard-sphere
potential, one indeed finds $g_2=0$ from Eq.\ (\ref{g2}).	 Our result is to be contrasted with the expression
$g_2=a^{2}-ar_{\rm e}/2$ based on the approximation (\ref{GaosDeltaE}) for
the interaction energy \cite{Gao03}.

As shown in Ref.\ \cite{Gao03}, the energy shift can be inserted
into the energy functional to obtain a generalization of the GP
equation for the condensate wave function $\Psi$ that contains a derivative term in the interaction energy:
\begin{equation}
i\hbar\frac{\partial}{\partial t}\Psi=\left[-\frac{\hbar^{2}}{2m}\nabla^{2}+V(r)+U_{0}\left( \left|\Psi\right|^{2}+g_{2}\nabla^{2}\left|\Psi\right|^{2}\right) \right]\Psi.
\end{equation}
Contributions from higher partial waves may be included through the addition of higher derivative terms, as described for Fermi systems in Ref.\ \cite{Roth01}.

\begin{figure}
\centerline{\includegraphics[width=8.6cm,clip=]{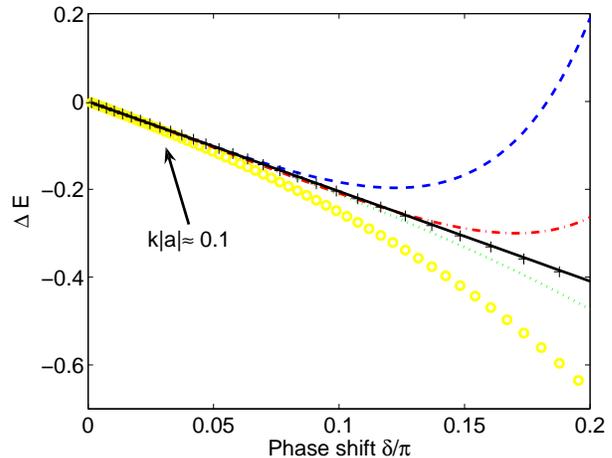}}
\caption{Energy shift for the ground state ($n=0$) of the combined potential discussed in the text.
The exact results (crosses) are compared with the energy shifts given by
the formula
$\Delta E=2 \pi \hbar^2 f|\psi(0)|^2/\mu$, where
$f=a$ (circles), $f=-\tan\delta/k$ (dotted line),
 $f=-\delta/k$ (solid line), by the result that includes the effective
range
 correction, $f=a(1-g_2k^2)$ (dashed-dotted line), and by the result of
 Ref.\ \cite{Gao03}, $f=a(1-\tilde{g}_2k^2)$ (dashed line). The energy is measured in
 units of $\hbar \omega$ and the core width is $L=0.25a_{\mathrm{ho}}$.
}
\label{fig:effRangeCorr}
\end{figure}

To test the validity of this prediction, we have performed numerical
integrations of the relative motion for the two-body problem considered
above, with a harmonic trapping potential and a short-range square well
potential, but for the lowest state rather than the excited states
considered in Sec.\ II.  The results are compared with those obtained
from analytical approximations.
Figure \ref{fig:effRangeCorr} shows that the inclusion of the
effective range correction with $g_2$ given by Eq.\ (\ref{g2}) greatly
improves the prediction given by the
simple scattering length approximation as soon as the condition $k|a|\ll
1$ is violated.

\section{Application to \mbox{Rydberg} molecules}
\label{sec:Rydb}

In \mbox{Rydberg} atoms the valence electron is in a highly excited
state with principal quantum number $n\gtrsim 20$, and within
quantum defect theory it is described in terms of generalized hydrogenic wave functions.
By analogy with the structures created around
positive ions in liquid helium \cite{Donnelly67} or BECs
\cite{Cote02,Massignan04}, a \mbox{Rydberg} atom with a large electric dipole
moment inside a BEC may create remarkable deformations of the condensate
density in its surroundings.  The additional degree of freedom
introduced by the permanent electric dipole moment could also be used to
realize conditional logic gates for quantum information processing
\cite{QuantInfProc}.

On the experimental side, \mbox{Rydberg} atoms have already been created in an
ultracold environment \cite{Grabowski05}, and there is a proposal to
excite and trap single \mbox{Rydberg} atoms inside a BEC \cite{Nogues}.  Being
overall electrically neutral, \mbox{Rydberg} atoms are not accelerated
by the stray electric fields which are unavoidable in
experimental vacuum chambers.  This is particularly relevant since, in a
typical apparatus for ultracold atoms, stray fields would drag an ion
outside the condensate in a time much less than 1 ms, making difficult
the observation of the induced density disturbances.


In recent work \cite{Greene00,Hamilton02}, it has been predicted that
the tailored excitation of single atoms in a BEC towards a \mbox{Rydberg} state
would induce the formation of molecules characterized by ultra-long
ranges ($R\sim2000$ a.u.) and very large permanent electric dipole
moments.  In these papers, the s-wave molecular potential between a ground
state atom and a \mbox{Rydberg} atom was taken to be

\begin{equation}
\label{GreenesPotential}
V_s(\vec{r},\vec{R})=
-\frac{2\pi \hbar^2 }{\mu}\frac{\tan\delta_0[k(R)]}{k}\delta(\vec{r}-\vec{R}),
\end{equation}
where $\vec{r}$ and $\vec{R}$ are the positions of the
electron and of the ground state atom relative to the \mbox{Rydberg} ion,
$\delta_0[k(R)]$ is the energy-dependent phase shift and the electron
wave number $k(R)$ is given by the hydrogenic relation
$k^2(R)/2-1/R=-1/2n^2$.  The authors of \cite{Greene00,Hamilton02}
follow Omont \cite{Omont77} and conclude that the appropriate potential for
the excited electron is given, in the Born-Oppenheimer approximation, by
\begin{equation}
\label{sfermi}
U_s(\vec R)=E_{nl}-\frac{2\pi \hbar^2 }{\mu}\frac{\tan\delta_0[k(R)]}{k} |\psi_{nl0}(\vec R)|^2,
\end{equation}
where $E_{nl}$ and $\psi_{nl0}$ are the unperturbed atomic
\mbox{Rydberg} energy and wave function (with quantum numbers $n\sim
30$, $l\lesssim 2$ and $m=0$). As discussed above, while the potential
in Eq.\ (\ref{GreenesPotential}) reproduces accurately eigenenergies and
wave functions of the Schr\"odinger equation, it should
not be used in mean-field calculations, where it yields unphysical
divergent energy shifts in the presence of a resonance.

This issue is particularly relevant for the scattering in the p-wave
channel, where the e-Rb(5s) scattering phase shift $\delta_{l=1} $ has a
resonance at an energy of approximately 30 meV, corresponding to a
distance between the Rb ion and the ground state atom of about 700 a.u.:
here Omont's expression for the  energy shift $\Delta E(\vec R)\propto
|\vec
\nabla
\psi_{nl0}(\vec R)|^2 \tan \delta_1(k)/k^3 $ diverges.  However,
according to the arguments we have given, the appropriate effective
interaction for a mean-field calculation is not given by this
expression, but rather by one with the tangent of the phase shift
replaced by the phase shift itself, and consequently there are no
divergence problems.

\section{Conclusion}
\label{sec:conc}

In this paper we have studied the expression for the effective two-body
interaction to be used in mean-field calculations of the energy of
a state.  All results agree in the limit of zero energy, but there are
differences at higher energies.  We have argued that the appropriate
effective interaction is proportional to the phase shift, rather than
other expressions that have been suggested, and we show that this holds
analytically for the states of two particles in a trap in the
quasiclassical approximation.
We have calculated energy
levels numerically for the problem of two particles interacting via a
short-range square-well potential in the presence of
a harmonic confining potential and have demonstrated that the analytical
expression in terms of the phase shift agrees well with the numerical
data, both for the ground state and for excited states with many nodes.
Since an effective interaction proportional to $\tan \delta$ gives
the correct wave functions and energy eigenvalues when inserted in
the Schr\"odinger equation, our results show that the choice of
effective interaction depends on the application.

In this article, we have also considered corrections to the
Gross--Pitaevskii equation to allow for the nonzero energy of the
relative motion of two particles and have derived a generalized
Gross--Pitaevskii equation that takes into account the effective range.
This equation gives a better approximation to the numerical results for
energy eigenvalues than does an earlier proposal \cite{Gao03}. A problem for future work is to include contributions
from higher partial waves in the GP equation. Finally,
we have argued that the effective two-body interaction to be used in
calculations of \mbox{Rydberg} atoms and molecules should be taken to be
proportional to the phase shift.

\begin{acknowledgments}
We are grateful to Alex Lande, J\"org Helge M\"uller, Tilman Pfau
and Mark Saffman for useful discussions.
\end{acknowledgments}


\begin{thebibliography}{1}
\bibitem{harmonic} For two particles each subject to a harmonic
trapping potential which is the same for both particles, the
center-of-mass and relative motions are separable.
\bibitem{breakdown}The effective range expansion Eq.~(\ref{effectiverange}) is valid only for potentials that fall off
more rapidly than $r^{-5}$ for $r\rightarrow \infty$. For example, in
Ref.\ \cite{O'Malley61} it was shown
that a potential with a $1/r^{4}$
tail the expansion of the s-wave phase shift contains a
term linear in $k$,
\begin{equation*}
k\cot\delta=-\frac{1}{a}+\frac{\pi\beta_{4}^{2}}{3a^{2}}k+Ak^{2}\ln\left(k\beta_{4}\right)+Bk^{2}+o\left(k^{2}\right),
\end{equation*}
where $\beta_{4}$, $A$ and $B$ are constants with dimensions
of a length.
\end{thebibliography}
\end{document}